\documentclass[12pt,preprint]{aastex}

\shorttitle{Structure of the Fe K-Edge}
\shortauthors{Palmeri et al.}
\usepackage[]{amsmath}
\begin{document}

\title{On the Structure of the Iron K-Edge}

\author{P. Palmeri\altaffilmark{1}, C. Mendoza\altaffilmark{2}, T. R. Kallman,}
\affil{NASA Goddard Space Flight Center, Greenbelt, MD 20771}
\and
\author{M. A. Bautista}
\affil{Centro de F\'{\i}sica, IVIC, Caracas 1020A, Venezuela}

\altaffiltext{1}{Research Associate, Department of Astronomy, University
of Maryland, College Park, MD 20742}
\altaffiltext{2}{Permanent address: Centro de F\'{\i}sica, IVIC, Caracas 1020A}

%------------------------------------------------------------------------------
\begin{abstract}
It is shown that the commonly held view of a sharp Fe K edge must
be modified if the decay pathways of the series of resonances
converging to the K thresholds are adequately taken into account.
These resonances display damped Lorentzian profiles of nearly
constant widths that are rapidly smeared to impose continuity
through the threshold. By modeling the effects of K damping on
opacities, it is found that the broadening of the K edge grows
with the ionization level of the plasma and that the appearance at
high ionization of a localized absorption feature at 7.2 keV is
identified as arising from the K$\beta$ unresolved transition
array.
\end{abstract}

\keywords{atomic processes---line formation---X-rays:
spectroscopy}

%------------------------------------------------------------------------------
\section{Introduction}

X-ray absorption and emission features arising from iron K-shell
processes are of practical importance in high-energy
astrophysics. This is primarily due to the iron cosmic abundance,
but also to the relatively unconfused spectral region where they
ubiquitously appear. Although the observational technology in
X-ray astronomy is still evolving, many of such features are
being resolved and consequently exploited in plasma diagnostics.
In this respect, they are naturally grouped according to their
origin, i.e. bound--bound or bound--free ionic transitions, and
much of the interpretation of the latter has relied on atomic
calculations
\citep{verner95,berrington97,donnelly00,berrington01} that
predict a sharp increase of the photoabsorption cross section at
the K-shell threshold. The purpose of this communication is to
emphasize that this commonly held view is incorrect due to an
oversimplified treatment of the decay pathways of the resonances
converging to this limit, and that previous astrophysical
inferences from K-edge structures should thus be revised.

%------------------------------------------------------------------------------
\section{Constancy of K damping}
When a photon is sufficiently energetic to promote a K-shell
electron to an excited Rydberg state, the latter decays through
radiative and autoionization (Auger) transitions. Illustrating
these processes in the relatively simple case of Ne-like Fe~{\sc
xvii}, the photoexcited K-vacancy states
\begin{equation}
h\nu+{\rm 1s}^2{\rm 2s}^2{\rm 2p}^6\longrightarrow
{\rm 1s2s}^2{\rm 2p}^6n{\rm p}
\end{equation}
have access to the following decay tree:
\begin{eqnarray}
{\rm 1s2s}^2{\rm 2p}^6n{\rm p} & \stackrel{{\rm K}n}{\longrightarrow}
         \label{beta}       & {\rm 1s}^2{\rm 2s}^2{\rm 2p}^6+h\nu_n \\
         \label{alpha}       & \stackrel{{\rm K}\alpha}{\longrightarrow} &
                  {\rm 1s}^2{\rm 2s}^2{\rm 2p}^5n{\rm p}+h\nu_\alpha \\
         \label{part}     & \stackrel{{\rm KL}n}{\longrightarrow} &
                          \begin{cases}
                  {\rm 1s}^2{\rm 2s}^2{\rm 2p}^5+e^- \\
                  {\rm 1s}^2{\rm 2s}{\rm 2p}^6+e^-
                          \end{cases}  \\
         \label{spec}     & \stackrel{{\rm KLL}}{\longrightarrow} &
                          \begin{cases}
                  {\rm 1s}^2{\rm 2s}^2{\rm 2p}^4n{\rm p}+e^- \\
                  {\rm 1s}^2{\rm 2s}{\rm 2p}^5n{\rm p}+e^- \\
                  {\rm 1s}^2{\rm 2p}^6n{\rm p}+e^-
                          \end{cases}
\end{eqnarray}
The radiative branches are controlled, as indicated in
Eqs.~(\ref{beta}--\ref{alpha}), by the K$\alpha$ ($2{\rm
p}\rightarrow {\rm 1s}$) array at $\lambda\sim 1.93$ \AA\ and the
K$n$ ($n{\rm p}\rightarrow {\rm 1s})$ of which the most salient
is the K$\beta\equiv{\rm K}3$ array at $\lambda\sim 1.72$ \AA.
C. Mendoza, P. Palmeri, T. Kallman, \& M. Bautista (2002, in preparation,
hereafter MPKB) have recently demonstrated that, for any
K-vacancy fine-structure state in the Fe isonuclear sequence, the
width ratio $\Gamma(\beta):\Gamma(\alpha)\lesssim 0.23$.

Equation~(\ref{part}) contains the participator Auger (KL$n$)
channels where the $n$p outer electron is directly involved in
the decay. By contrast, in the KLL channels (Eq.~\ref{spec}) the
$n$p Rydberg electron remains a spectator. It has also been
demonstrated by MPKB that, for any Fe K-vacancy
fine-structure state with a filled L shell, the total K$\alpha$
and KLL widths are practically independent of both the principal
quantum number $n\geq 3$ of the outer-electron configuration and
the electron occupancy $N>9$, and keep a ratio of $\Gamma({\rm
KLL}):\Gamma(\alpha)\sim 1.5$. These findings illustrate the
works of Gauss's law at the atomic scale as previously reasoned by
\citet{manson91} regarding the inner-shell properties of Kr and
Sn. Furthermore, MPKB have also determined that Auger
decay is 94\% dominated by the KLL channels in Fe~{\sc xvii} and
by no less than 70\% in Fe~{\sc x}. Since both the K$n$ and KL$n$
widths decay with the effective quantum number as
$\sim(n^{*})^{-3}$, the total widths of K-vacancy states within
an isonuclear sequence may certainly be assumed constant for
$n\rightarrow \infty$ for all members with $N>3$. In line with
the previous discussions by \citet{gorczyca00a} and
\citet{gorczyca00b}, the width constancy of high-$n$ K resonances
within an isonuclear sequence for members with $N>3$ procure
indelible effects on the threshold structures of their
photoabsorption cross sections which have indeed been measured in
the laboratory in O~{\sc i} and Ne~{\sc i} but seriously
underestimated in many atomic calculations.

%-----------------------------------------------------------------------------
\section{The Fe~{\sc xvii} case}

We make use of the Breit--Pauli R-matrix ({\sc bprm)} method
\citep{berrington78} to study K damping in Fe~{\sc xvii}. This
numerical approach is based on the close-coupling approximation
whereby the wavefunctions for states of an $N$-electron target
and a colliding electron are expanded in terms of a finite number
of target eigenfunctions. As can be deduced from Eq. (\ref{spec}),
{\sc bprm} calculations of photoionization or electron impact
excitation cross sections can handle the spectator Auger channels
only for low-$n$ K resonances since the close-coupling expansion
must explicitly include $nl$ target states. However,
\citet{gorczyca99} has modified {\sc bprm} to implicitly account
for the spectator Auger channels by means of an optical potential
devised from multichannel quantum defect theory. A target-state
energy now acquires the imaginary component
\begin{equation}
E_i\to E_i-i\Gamma_i/2
\end{equation}
where $\Gamma_i$ is its partial decay width. This treatment of spectator
Auger decay
is thus analogous to that of radiation damping \citep{damping}.

In order to discern the contributions of the KLL channels to the
total Auger widths, several {\sc bprm} calculations are carried
out for the ${\rm Fe}^{17+}+e^{-}$ system. Firstly, only target
levels from the ${\rm 1s}^2{\rm 2s}^2{\rm 2p}^5$, ${\rm 1s}^2{\rm
2s}{\rm 2p}^6$, and ${\rm 1s}{\rm 2s}^2{\rm 2p}^6$ configurations
are included in the expansion thus only accounting for
participator Auger decay. In a second calculation, levels from
the ${\rm 1s}^2{\rm 2s}^2{\rm 2p}^4n{\rm p}$, ${\rm 1s}^2{\rm
2s}{\rm 2p}^5n{\rm p}$, ${\rm 1s}^2{\rm 2p}^6n{\rm p}$
configurations with $n\leq 4$ are added to the target
representation which now maps out the complete Auger manifold for
the ${\rm 1s}^{-1}{\rm 3p}$ and ${\rm 1s}^{-1}{\rm 4p}$
resonances. In Table~\ref{widths}, the resulting Auger widths for
these resonances are compared, where increases by KLL decay
greater than an order of magnitude and constant total Auger
widths are depicted.

\begin{figure}
\epsscale{0.9}
\plotone{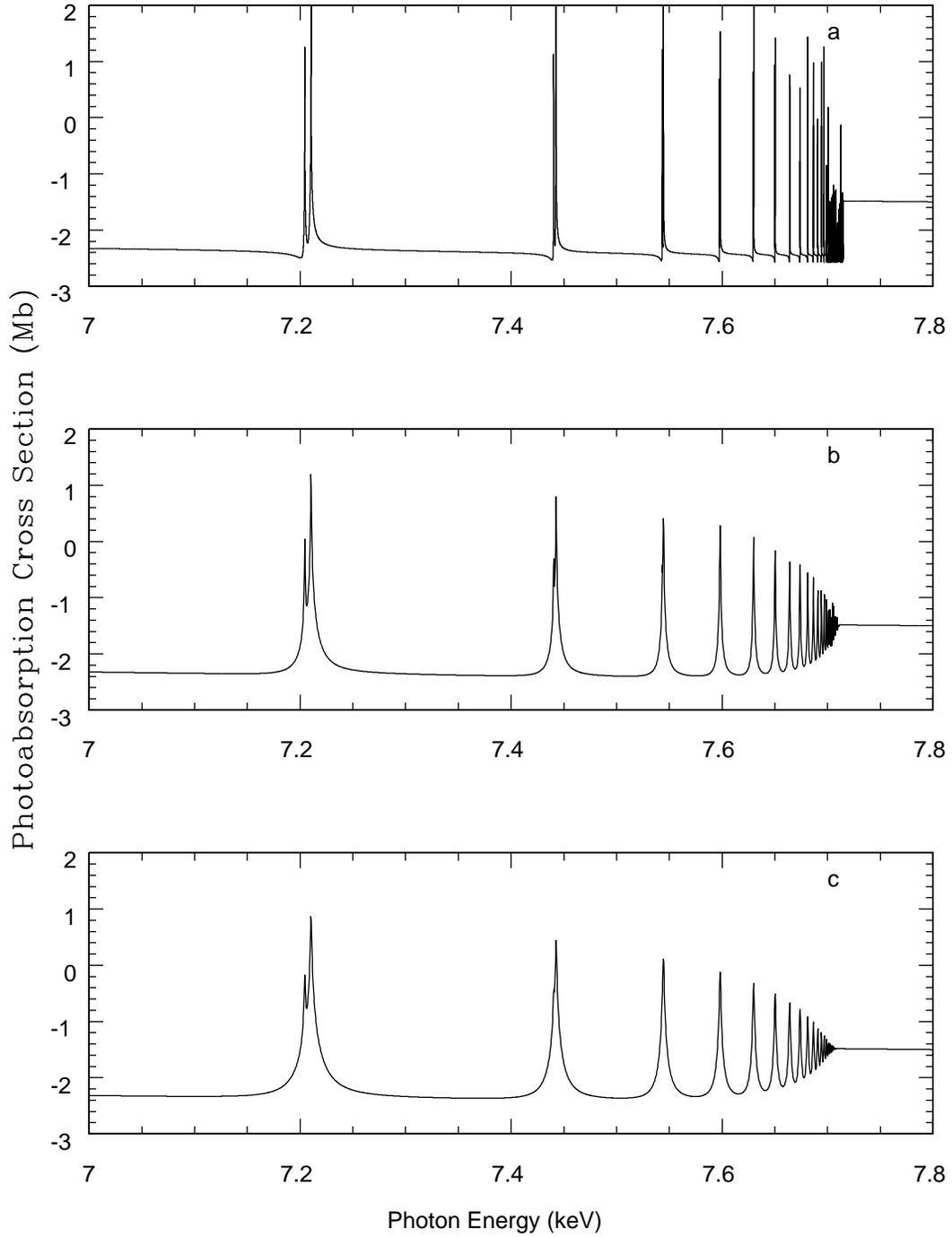}
\caption{Total photoabsorption
cross section of Fe~{\sc xvii} computed with {\sc bprm} (a) without
damping, (b) with radiation damping, and (c)
with both radiation and spectator Auger damping.
\label{fig1}}
\end{figure}

In a further calculation, the {\sc bprm}+optical potential
approach of \citet{gorczyca99} is employed to study the effects
of radiation and Auger damping on the resonance structure.
Fig.~\ref{fig1} shows the photoabsorption cross section of the
${\rm 1s}^2{\rm 2s}^2{\rm 2p}^6\ ^2{\rm S}_{1/2}$ ground state of
Fe~{\sc xvii} in the near K-threshold region. It may be seen that
the undamped cross section is populated by a double series (${\rm
1s}^{-1}n{\rm p}\ ^{3,1}{\rm P}^{\rm o}_1$) of narrow and
asymmetric resonances that converge to a sharp K edge. The
inclusion of radiation (K$\alpha$) and Auger (KLL) dampings leads
to resonance series with constant widths and symmetic profiles
that get progressively smeared with $n$ to produce a smooth
transition through the K threshold; in low resolution, the edge
would thus appear to be downshifted in energy. The symmetric
profiles are, as discussed by \citet{nayandin01}, a consequence
of the fact that KLL channels give rise to continuum states that can
only be reached by dipole forbidden direct photoionization of
the ground state and thus cause the Fano $q$-parameter to tend toward infinity.
With respect to radiation damping, the dominance of
the K$\alpha$ over the Auger participator channels causes the modified Fano 
$q$-parameter \citep{damping} to be large, so as to also yield 
a symmetric profile.
Resonance smearing is the result of oscillator
strength conservation which must enforce an approximate
$(n^{*})^{-3}$ reduction of the resonance peak-values due to
their constant widths.

%-------------------------------------------------------------------------------

\section{K-edge spectral signatures}
An advantage of the simple behavior of K damping in Fe is that
its impact on the opacities can be estimated with an analytic
model. The latter has been established and validated with the
{\sc bprm} photoabsorption cross section of Fe~{\sc xvii}: the
damped ${\rm 1s}^{-1}3{\rm p}$ resonances are fitted with
Lorentzian profiles and extrapolated with $n^*$ assuming constant
widths and intensities decreasing with $(n^{*})^{-3}$, the
resonance positions being deduced by the usual Ritz formula. The
cross sections of all the other Fe ions in the near threshold
region are then determined in a similar fashion assuming a single
${\rm 1s}\rightarrow n{\rm p}$ Rydberg series of smeared
resonances for each species with $n\geq 2$ for $4\leq N\leq 9$,
$n\geq 3$ for $10\leq N\leq 17$, and $n\geq 4$ for $18\leq N\leq
26$. Since the resonance widths can be assumed constant with $N$
for $N>9$ (MPKB), they are assigned the value of the
3p resonances in Fe~{\sc xvii}, namely, $4.5\times 10^{-2}$ ryd. For
$N\leq 9$, this value is scaled by a factor of $N(N-1)/90$. The
resonance intensities of the first few members of each series are
obtained from multiplet $f$-values computed with the atomic
structure code {\sc hfr} \citep{cowan81}, and the K-threshold
positions and background cross sections are taken from
\citet{verner95}.

\begin{figure}
\epsscale{0.8}
\plotone{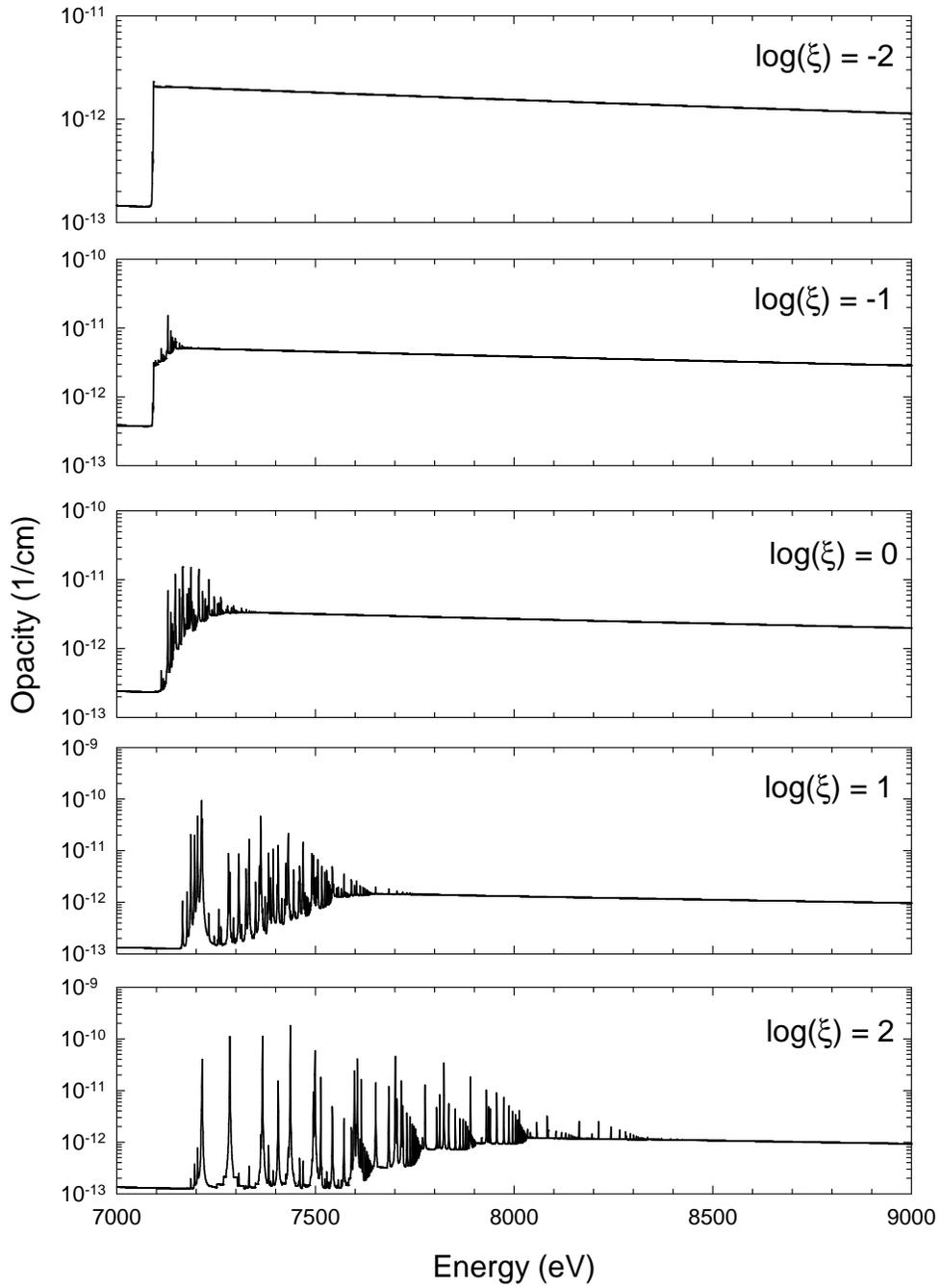}
\caption{Opacities of a
photoionized gas with solar elemental abundances as a function of
photon energy in the region of the iron K edge.  Plots correspond
to different values of the ionization parameter $\xi$.
\label{fig2}}
\end{figure}

The monochromatic opacities are generated with the {\sc xstar}
program \citep{kb01}; the self-consistent ionization balance and
electron temperature are computed under the assumption that
ionization and heating are primarily due to an external source of
continuum photons and that all processes are in a steady state.
Solar abundances and a $F_\epsilon\sim \epsilon^{-1}$ continuum
power law are adopted. As shown in Fig.~\ref{fig2}, results are
characterized in terms of the familiar ionization parameter
$\xi\equiv L/nR^2$ in the range $0.01\leq \xi\leq 100$ where $L$
is the luminosity of the incident X-ray radiation, $n$ the gas
density, and $R$ the distance from the radiation source. Two
distinctive spectral signatures emerge: (i) the concept of a sharp
edge only applies for lowly ionized ($\log \xi\sim -2$) plasmas,
and its progressive broadening is a function of ionization; (ii)
for $\log \xi\gtrsim 0$, a strong absorption feature appears at
the piedmont ($\sim 7.2$ keV) caused by the K$\beta$ unresolved
transition array (UTA).

\begin{figure}
\epsscale{1.}
\plotone{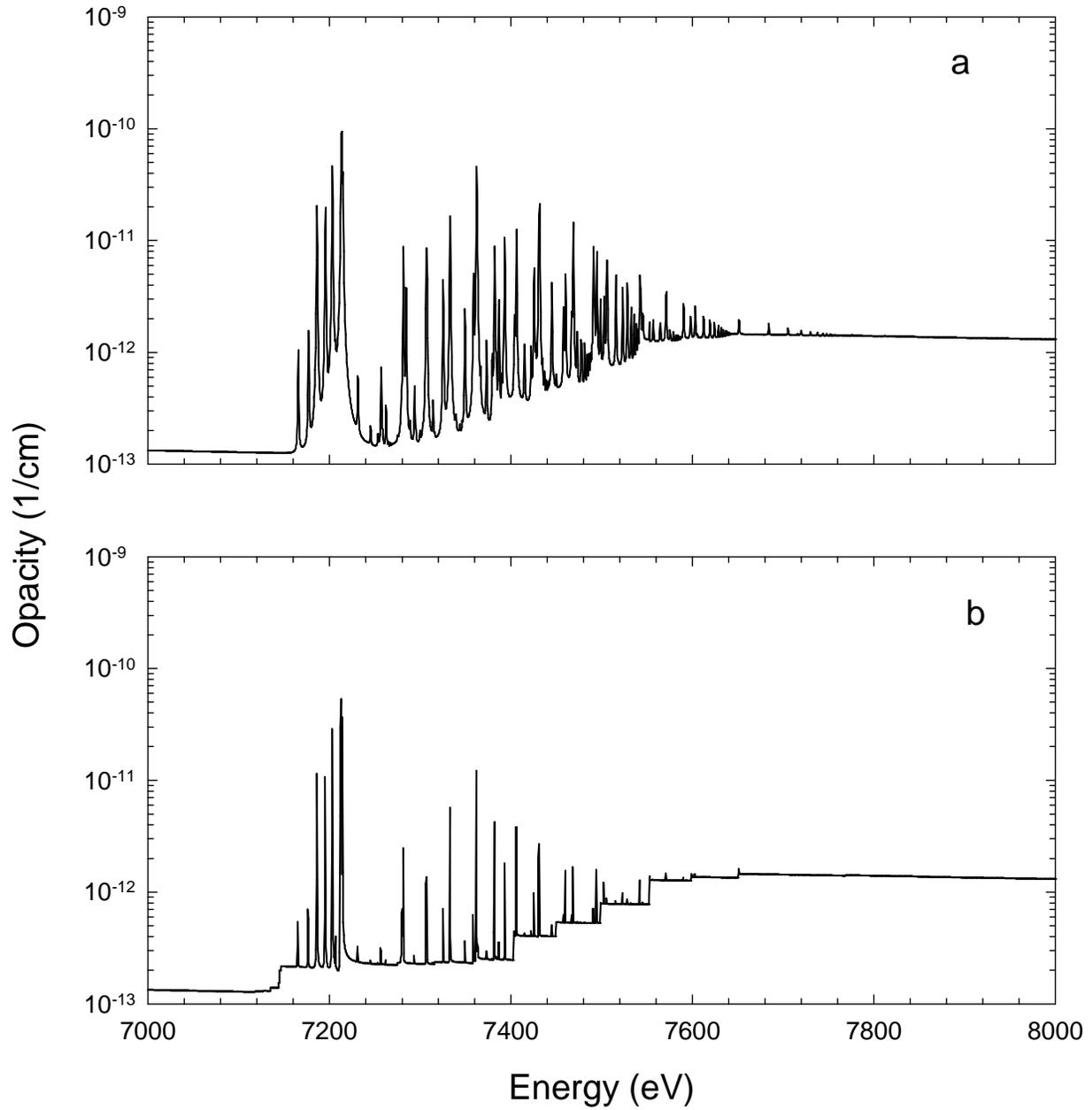}
\caption{Opacities of a
photoionized gas with solar elemental abundances and an ionization parameter $\xi = 10$, including (a) K damped photoionization cross sections and
 (b) undamped cross sections.
\label{fig3}}
\end{figure}

The features produced by the smeared resonances on the K-edge opacities
really stand out in a comparison with opacities computed with undamped
photoionization cross sections. The latter can be approximated by
replacing the Lorentzian profiles with Fano type with
constant asymmetry parameters $q = 200$, i.e., the undamped value
for the 3p resonance in Fe~{\sc xvii}. The undamped widths are
obtained with {\sc hfr} for the first members of each series and
scaled down by $1/{n^{*}}^{3}$ for the higher. The resonant cross
section is deduced from the following approximate formula:
\begin{equation}
q^2 \approx {2 f(g,r) \over \pi \Gamma f(g,c)}
\label{resxs}
\end{equation}
where $f(g,r)$, calculated with {\sc hfr}, is the oscillator strength 
for the transition between the ground state and the noninteracting resonance,
$\Gamma$ is the resonance width, and $f(g,c)$ is the
oscillator strength for the transition between the ground and the
continuum states and is proportional to the resonant cross
section. A comparison between damped and undamped opacities at $\xi$ = 10
is shown in Fig.~\ref{fig3}, in which the steady upclimb in the former is
 replaced by sharp steps in the latter. Moreover, the role of resonances
 in the damped case, in particular the K$\beta$ UTA, becomes more dominant.
%-------------------------------------------------------------------------------
\section{Observations}

Broad K-edge absorption structures have been widely observed in
the X-ray spectra of active galactic nuclei and black-hole
candidates as reviewed by \citet{ebisawa94}and more recently
 reported by \citet{done99} and
\citet{miller02}. They have been interpreted in terms of the
reflection of X-rays by optically thick accretion disks around
central compact objects, partial absorption models, or
relativistic broadening. Furthermore, an unidentified local
absorption feature at $\sim 7$ keV just below a smeared K edge
has been reported by \citet{ebisawa94} in the bright 
X-ray nova GS 1124--68 and by \citet{pounds02} in the Seyfert 1
galaxy MCG-6-30-15; in this respect, the fitted ionization parameter 
($\xi \sim 5.9$) in the latter seems to support our predicted edge
signatures. It is expected that the present findings will
contribute to a more realistic spectral modeling above 7 keV
that has recently been described as critical by \citet{pounds02}.
However, quantitative modeling implies a revision of the
inner-shell photoionization cross sections for the Fe isonuclear
sequence which is currently underway.

%-------------------------------------------------------------------------------

\acknowledgments
We are indebted to an anonymous referee for suggestions that have improved
the presentation of the material in this report.
C. M. acknowledges a Senior Research Associateship from the National
Research Council.
M. A. B. acknowledges partial support from FONACIT, Venezuela, under Proyect No.
S1-20011000912.

%----------------REFERENCES-----------------------------------------------------

%--------------------------------Table 1---------------------------------------------------
\newpage
\begin{deluxetable}{lrrr}
\tablecaption{\label{widths} Auger widths for the 1s$^{-1}n$p resonances in
Fe~{\sc xvii}}
\tablewidth{0pt}
\tablehead{ \colhead{State}&\colhead{$E$}&
\colhead{$\Gamma({\rm KL}n)$\tablenotemark{a}}&
\colhead{$\Gamma({\rm KL}n+{\rm KLL})$\tablenotemark{b}}\\
\colhead{}&\colhead{(eV)}&\colhead{(eV)}&\colhead{(eV)}
} \startdata
1s$^{-1}$3p~$^3$P$^{\rm o}_1$ & 7.187 & 3.80$\times$10$^{-2}$ & 6.50$\times$10$^{-1}$ \\
1s$^{-1}$3p~$^1$P$^{\rm o}_1$ & 7.192 & 2.17$\times$10$^{-2}$ & 6.34$\times$10$^{-1}$ \\
1s$^{-1}$4p~$^3$P$^{\rm o}_1$ & 7.422 & 1.39$\times$10$^{-2}$ & 6.15$\times$10$^{-1}$ \\
1s$^{-1}$4p~$^1$P$^{\rm o}_1$ & 7.425 & 7.95$\times$10$^{-3}$ & 6.08$\times$10$^{-1}$ \\
\enddata
\tablenotetext{a}{Includes only participator (KL$n$) Auger channels}
\tablenotetext{b}{Includes both participator (KL$n$) and spectator (KLL) Auger channels}
\end{deluxetable}

%---------------------Fig. 1--------------------------------------------------
\newpage

%---------------------Fig. 2--------------------------------------------------
\newpage

%---------------------Fig. 3--------------------------------------------------
\newpage
%--------------------------------------------------------------------------------------------
%

\end{document}